\documentclass[sigplan,screen]{acmart}

\AtBeginDocument{%
  }

\setcopyright{acmcopyright}
\copyrightyear{2018}
\acmYear{2018}
\acmDOI{XXXXXXX.XXXXXXX}

\acmConference[Conference acronym 'XX]{Make sure to enter the correct
  conference title from your rights confirmation emai}{June 03--05,
  2018}{Woodstock, NY}
\acmPrice{15.00}
\acmISBN{978-1-4503-XXXX-X/18/06}

\copyrightyear{2024}
\acmYear{2024}
\setcopyright{rightsretained}
\acmConference[DaMoN '24]{20th International Workshop on Data Management on New Hardware }{June 10, 2024}{Santiago, AA, Chile}
\acmBooktitle{20th International Workshop on Data Management on New Hardware (DaMoN '24), June 10, 2024, Santiago, AA, Chile}
\acmDOI{10.1145/3662010.3663439}
\acmISBN{979-8-4007-0667-7/24/06}

\begin{document}

\title{SFVInt: Simple, Fast and Generic Variable-Length Integer Decoding using Bit Manipulation Instructions}
\author{Gang Liao$^{*\dagger}$ $\quad$ Ye Liu $\quad$ Yonghua Ding $\quad$ Le Cai $\quad$ Jianjun Chen}
\affiliation{%
  \institution{$^{*}$ByteDance Infrastructure System Lab $\quad$ $^{\dagger}$University of Maryland College Park}
  \country{}}
\affiliation{gangliao@umd.edu $\quad$ \{ye.liu, yonghua.ding, le.cai, jianjun.chen\}@bytedance.com\country{}}







\renewcommand{\shortauthors}{Gang Liao et al.}

\begin{abstract}
The ubiquity of variable-length integers in data storage and communication necessitates efficient decoding techniques. In this paper, we present SFVInt, a simple and fast approach to decode the prevalent Little Endian Base-128 (LEB128) varints. Our approach effectively utilizes the Bit Manipulation Instruction Set 2 (BMI2) in modern Intel and AMD processors, achieving significant performance improvement while maintaining simplicity and avoiding overengineering. 
SFVInt, with its generic design, effectively processes both 32-bit and 64-bit unsigned integers using a unified code template, marking a significant leap forward in varint decoding efficiency. We thoroughly evaluate SFVInt's performance across various datasets and scenarios, demonstrating that it achieves up to a 2x increase in decoding speed when compared to varint decoding methods used in established frameworks like Facebook Folly and Google Protobuf. 
\end{abstract}

\maketitle

\section{Introduction}
Variable-length integers (varints) play a crucial role in optimizing space efficiency for integer data representation across a wide range of systems and applications. From search engines like Apache Lucene~\cite{lucene} and databases such as IBM DB2~\cite{10.14778/1687553.1687573} and Apache Kudu~\cite{lipcon2015kudu}, to columnar formats like ORC~\cite{orc} and Parquet~\cite{parquet}, varints serve as a fundamental building block for compact data storage and processing. The importance of varints extends beyond traditional data management systems. They are extensively used in serialization frameworks such as Google Protobuf~\cite{protobuf} and the WebAssembly binary encoding~\cite{webassembly-binary}, highlighting their significance in cross-platform data exchange and web-based applications. Moreover, varints are an integral part of the Go programming language's default API~\cite{golang}, further emphasizing their widespread adoption across various critical technological domains~\cite{gang2023flock,bullfrog, filescale,gang_thesis}.

Among the various varint encoding schemes, the Little Endian Base 128 (LEB128)~\cite{leb128} has emerged as a widely adopted standard due to its optimal balance of space efficiency and implementation simplicity. LEB128 encodes integers as a sequence of bytes, where the most significant bit of each byte indicates whether more bytes follow, allowing for variable-length representation. However, despite its space-saving advantages, the decoding process for LEB128 varints often becomes a performance bottleneck in data-intensive applications. The unpredictable lengths of the encoded integers lead to branch mispredictions and limit the opportunities for vectorization, hindering the overall efficiency of the decoding process.

To address this challenge, we leverage the Bit Manipulation Instruction Set 2 (BMI2)~\cite{bmi2}, a powerful set of instructions available in modern Intel~\cite{intel64andia32v2b} and AMD~\cite{amd64archmanualvol3} CPUs. BMI2 offers advanced bit manipulation capabilities that can significantly accelerate the decoding of LEB128 varints. By harnessing these instructions, we introduce SFVInt, a simple, fast, and generic approach for varint decoding.
SFVInt stands out for its simplicity and performance. With just 500 lines of code, our solution achieves up to a 2x speedup in decoding compared to widely-used systems and libraries. SFVInt's generic design seamlessly handles both 32-bit and 64-bit unsigned integers using a unified C++ template, enhancing its versatility and maintainability.
Our in-depth analysis highlights the critical role of BMI2 instructions in achieving this substantial performance improvement. We demonstrate how SFVInt strategically employs these instructions to extract and manipulate the relevant bits from the encoded data, enabling efficient decoding while minimizing branch mispredictions.



To the best of our knowledge, our work represents the first comprehensive evaluation of BMI2 instructions in the context of varint decoding, marking a significant advancement in this field.
By leveraging the powerful bit manipulation capabilities offered by these instructions, we have achieved a critical enhancement in the efficiency of varint decoding.
The implications of our research extend beyond the scope of this paper, potentially influencing the design and performance optimization of a wide range of data-intensive systems. Databases, network protocols, and big data frameworks are just a few examples of the domains that can benefit from the improved varint decoding performance brought forth by our approach.


\section{Background}
\subsection{Variable-Length Integers}

Varints, or variable-length integers, are a method of encoding integers using one or more bytes. They are widely adopted in computing for their ability to optimize space usage. The principal idea behind varints is to use a dynamic number of bytes for representation, depending on the integer's magnitude. Smaller numbers consume fewer bytes, making this method highly space-efficient for data that predominantly consists of smaller integers.

Varint encoding, essential in data format optimization, encompasses several schemes, including Group Varint~\cite{jeff37797}, Prefix Varint~\cite{jeff37797}, and Stream Varint~\cite{Lemire_2018}, each with its unique characteristics. Among these, LEB128 stands out for its simplicity and widespread adoption. LEB128 operates on a simple yet effective principle: it encodes an integer into a sequence of bytes, where each byte holds 7 bits of the integer plus a continuation bit. This continuation bit indicates the presence of subsequent bytes. In this scheme, the integer's least significant 7-bit group is positioned in the first byte, adhering to the little-endian convention. Despite the existence of more complex varint formats, LEB128's simplicity makes it a prevalent choice in various applications, underlining its effectiveness in efficiently handling integer encoding.

The decoding process of LEB128 involves reading each byte, extracting the 7 bits of the integer, and assembling them to reconstruct the original integer. This process continues byte by byte until a byte with the continuation bit set to 0 is encountered, indicating the end of the encoded integer.

LEB128 decoding, a bottleneck in high-volume data processing, faces challenges due to encoded integers' unpredictable lengths. This variability frequently leads to branch mispredictions, exacerbating CPU overhead. The need for per-byte decision-making during decoding amplifies this issue, especially in large-scale or real-time processing environments. Addressing these inefficiencies is crucial for optimizing LEB128's decoding performance.

\subsection{Bit Manipulation Instructions (BMI)}

Bit Manipulation Instructions (BMI2)~\cite{bmi2} are advanced CPU instruction sets designed for efficient bit-level operations, introduced with Intel's Haswell architecture and adopted by AMD. These instructions, including \texttt{PDEP} and \texttt{PEXT}, enable faster and more effective manipulation of bits in registers or memory. BMI2's inclusion in modern processors optimizes critical operations in high-performance computing, data processing, and cryptographic applications, reflecting a broader trend towards enhancing computational efficiency through specialized hardware capabilities.

\begin{figure}[h!]
\centering
\includegraphics[width=0.4\textwidth]{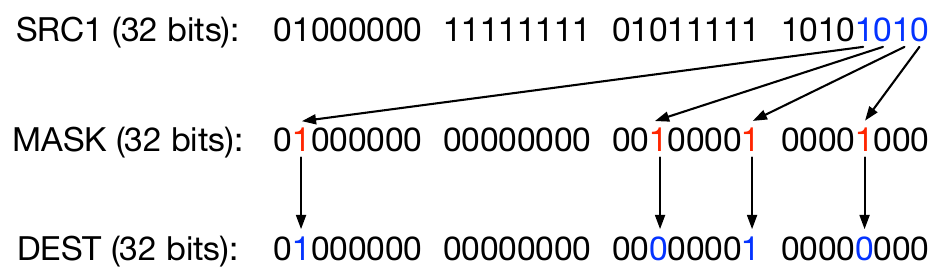}
\vspace{-3mm}
\caption{\texttt{PDEP} Example.}
\vspace{-3mm}
\label{pdep}
\end{figure}

\texttt{PDEP} (Parallel Bit Deposit): This instruction streamlines the process of depositing bits from a source operand into a destination operand based on a mask. It allows for selectively placing bits in specific positions, enabling precise control over bit placement. \texttt{PDEP} is particularly useful in scenarios where bits from a variable need to be reorganized or masked in a specific pattern, as illustrated in Figure \ref{pdep}.

\begin{figure}[h!]
\centering
\includegraphics[width=0.4\textwidth]{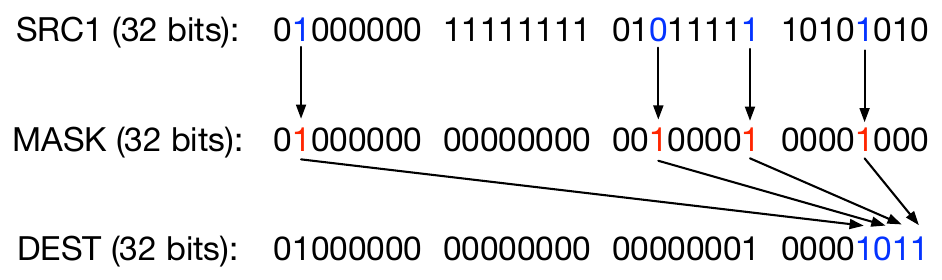}
\vspace{-3mm}
\caption{\texttt{PEXT} Example.}
\vspace{-3mm}
\label{pext}
\end{figure}

\texttt{PEXT} (Parallel Bit Extract): \texttt{PEXT} complements \texttt{PDEP} by enabling the extraction of bits from a source operand based on a mask. This instruction allows for the selective extraction of bits, effectively 'filtering' the source operand through the mask to produce the desired bit pattern in the destination operand. \texttt{PEXT} is invaluable for operations that require isolating specific bits from a larger set, such as decoding varints where bits need to be extracted from a sequence of bytes. An example of \texttt{PEXT} in action is depicted in Figure \ref{pext}.

Our motivation for utilizing BMI2 in LEB128 varint decoding lies in its ability to simplify the bit extraction and assembly process, crucial for efficient decoding. Traditional LEB128 decoding, characterized by sequential bitwise operations, benefits from the parallelism and efficiency offered by BMI2 instructions.

\section{SIMD Implementation}

This section outlines traditional varint encoding/decoding methods, then focuses on optimizing operations like varint sequence skipping and storage size determination. We explore BMI2-enhanced implementation, detailing how it improves decoding efficiency and performance, demonstrating a progression from basic to advanced techniques in varint processing.

\subsection{Basic Varint Operations}
\label{basic_varint}
\textbf{Varint Encoding}: LEB128 encodes integers into a compact, variable-length format by dividing each integer into 7-bit blocks, each represented by a single byte. All bytes, barring the final one, mark their most significant bit as 1 to denote the sequence's continuation. The final byte's most significant bit is set to 0, indicating the end of the encoding. This method is efficient for storing integers, particularly smaller values. The pseudocode below details this encoding process:

{\footnotesize
\begin{verbatim}
Algorithm 1: LEB128 Integer Encoding
Input: Integer val, BufferPointer buf, Output: None
1. while val >= 0x80 do
2.   *buf = 0x80 | (val & 0x7f);
3.   val >>= 7; buf++;
4. end while
5. *buf = val; buf++;
\end{verbatim}
}

If \texttt{val} is 128 (0x80) or greater, the function stores the lowest 7 bits of \texttt{val} into the \texttt{buf}, setting the 8th bit of this byte (denoted by 0x80) to signal that more bytes follow (line 2).
The buffer pointer \texttt{buf} is then advanced, and \texttt{val} is right-shifted by 7 bits to prepare the next 7 bits for encoding (line 3).
This process repeats until \texttt{val} is less than 128. At this point, \texttt{val} is small enough to fit into the final byte, indicating the end of the variable-length integer (line 5). 

\vspace{2mm}
\noindent \textbf{Varint Decoding}: LEB128 extracts the original integer from its encoded representation. It involves iteratively processing each byte, reconstructing the integer by aggregating 7-bit segments.

{\footnotesize
\begin{verbatim}
Algorithm 2: LEB128 Integer Decoding (Basic Version)
Input: BufferPointer buf
Output: T res (either uint32_t or uint64_t)
1. res = 0; max_shift = std::is_same_v<T, uint32_t> ? 28 : 63;
2. for shift = 0 to max_shift step 7 do
3.   res |= ((*buf & 0x7f) << shift);
4.   if (likely(!(*buf++ & 0x80))) break;
5. end for
\end{verbatim}
}

In the basic LEB128 decoding algorithm, each byte is read and its lower 7 bits are combined into the result integer using bitwise operations. The most significant bit of each byte serves as a continuation flag, indicating whether more bytes follow. The decoding process concludes when a byte with its most significant bit unset is encountered, signaling the end of the varint.

The maximum shift values for decoding are determined based on the target integer size (line 1). For 32-bit integers, the maximum shift is 28, while for 64-bit integers, it is 63. These values are derived from the fact that LEB128 uses a 7-bit encoding scheme. To illustrate this, consider a 32-bit unsigned integer with a maximum value of 4,294,967,295 ($2^{32} - 1$). Representing this value using LEB128 would require up to 5 bytes, as $\lfloor 32 / 7 \rfloor = 4$, with 4 bits remaining. The shift values used in the decoding process correspond to each 7-bit group in these five bytes: 0, 7, 14, 21, and 28 (line 2).

This basic decoding algorithm serves as a foundation for understanding LEB128 decoding, but it may not be optimal for performance due to the byte-by-byte processing and the presence of conditional branches. More advanced techniques, such as vectorization and branch elimination, can be applied to improve decoding efficiency.


\vspace{2mm}
\noindent \textbf{Varint Skipping}: Efficient skip operations are crucial for effective filtering and scanning of varint-encoded data streams in columnar storage systems. By leveraging SIMD instructions and bitwise operations, we employ a technique that processes 64-bit data blocks, enabling simultaneous handling of multiple varints, improving performance compared to byte-by-byte approaches.

{\footnotesize
\begin{verbatim}
Algorithm 3: LEB128 Integer Skipping
Input: uint64_t n (# integers to skip), BufferPointer buf
Output: Updated BufferPointer
1. res = 0; w = (const uint64_t*)(buf);
2. while n >= 8 do
3.   n -= __builtin_popcountll(~(*w++) & 0x8080808080808080);
4. end while
5. buf = (const char*)(w);
6. while n-- do
7.   while *buf++ & 0x80 do {;} end while // fallback
8. end while
\end{verbatim}
}
\vspace{-1mm}

The algorithm above takes two input parameters: \texttt{n}, the number of integers to skip, and \texttt{buf}, a pointer to the buffer containing the encoded integers. It initializes a 64-bit pointer \texttt{w} to the same location as \texttt{buf}, enabling the processing of 64-bit words (line 1). The condition \texttt{n >= 8} in line 2 ensures that the algorithm processes the buffer in 64-bit word-sized chunks as long as there are at least 8 integers to skip. This threshold is chosen because a 64-bit word can contain up to 8 bytes, each potentially representing a single-byte varint.

Within the loop, the algorithm performs a series of bitwise operations to determine the number of complete varints in the current 64-bit word (line 3). It first inverts all the bits of the dereferenced 64-bit word using the bitwise NOT operation ($\sim$). For LEB128 encoding, the most significant bit (MSB) of each byte being '1' indicates the continuation of a varint, and '0' indicates its termination. By inverting these bits, the continuation bits ('1's) are turned into '0's and vice versa. Then, it applies a mask (\texttt{0x8080808080808080}) to isolate the MSBs of each byte, which indicate the continuation or termination of a varint. By performing the bitwise AND operation ($\&$) between the inverted word and the mask, the algorithm obtains a 64-bit value where the most significant bits of each byte are set to 1 if the corresponding varint terminates and 0 if it continues. 

The \texttt{POPCNT} function is then used to efficiently count the number of set bits (varint terminations) in the resulting 64-bit value. This count represents the number of complete varints in the current word. The algorithm subtracts this count from \texttt{n} to update the number of remaining integers to skip (line 3).
After the loop, the algorithm updates the \texttt{buf} pointer to the location where the 64-bit word processing stopped (line 5). It then enters a byte-by-byte skipping loop (lines 6-8) to handle any remaining integers. This loop continues until \texttt{n} reaches zero, indicating that the desired number of integers has been skipped.

By processing the buffer in 64-bit words and using efficient bitwise operations and SIMD instructions, the LEB128 Integer Skipping algorithm achieves fast skipping of varints, making it suitable for scenarios where large numbers of integers need to be skipped quickly.



\vspace{2mm}
\noindent \textbf{Varint Sizing}: In varint-based systems, accurately estimating storage needs is crucial for efficient memory allocation and management. We utilize a precomputed lookup table (LUT) to quickly calculate the exact byte size needed for varint encoding, based on the position of the most significant set bit in integers. This approach optimizes computational efficiency with low overhead in caching, facilitating rapid estimations for dynamic storage allocation. 
{\footnotesize
\begin{verbatim}
Algorithm 4: LEB128 Integer Sizing
Input: vec<T> values (T is either uint32_t or uint64_t)
Output: uint64_t size
1. kLUT = {10, 
      9, 9, 9, 9, 9, 9, 9, 8, 8, 8, 8, 8, 8, 8,
      7, 7, 7, 7, 7, 7, 7, 6, 6, 6, 6, 6, 6, 6,
      5, 5, 5, 5, 5, 5, 5, 4, 4, 4, 4, 4, 4, 4,
      3, 3, 3, 3, 3, 3, 3, 2, 2, 2, 2, 2, 2, 2,
      1, 1, 1, 1, 1, 1, 1};
3. size = 0;
4. for each val in values do
5.   if constexpr (std::is_same_v<val, uint32_t>) then
6.       size += kLUT[__builtin_clz  (val | 1) + 32]
7.   else
8.       size += kLUT[__builtin_clzll(val | 1)]
9. end for
\end{verbatim}
}

The algorithm computes the total byte size required for varint encoding a collection of integers. It leverages a precomputed LUT that maps the position of the most significant set bit in an integer to the corresponding varint byte size. The LUT is constructed based on the observation that the byte size needed for varint encoding depends on the magnitude of the integer value.

For each integer \texttt{val} in the input vector \texttt{values}, the algorithm determines the position of the most significant set bit using the LZCNT (Leading Zero Count) instruction, depending on the integer type (\texttt{uint32\_t} or \texttt{uint64\_t}). To handle the case where val is zero, the algorithm performs a bitwise \texttt{OR} operation with 1 \texttt{(val|1)} before counting the leading zeros\footnote{It's worth noting that BSF and BSR, with a source operand value of (0), leave the destination undefined and set the ZF (zero flag). However, C++20's \texttt{std::countl\_zero} doesn't have this issue, if your compiler supports it.}. This ensures that the LUT index is correctly calculated even for zero values.
The position of the most significant set bit is then used as an index to retrieve the corresponding byte size from the LUT. For \texttt{uint32\_t} values, an offset of 32 is added to the LUT index to account for the difference in bit positions between 32-bit and 64-bit integers.

\begin{figure}
    \centering
    \includegraphics[width=1\linewidth]{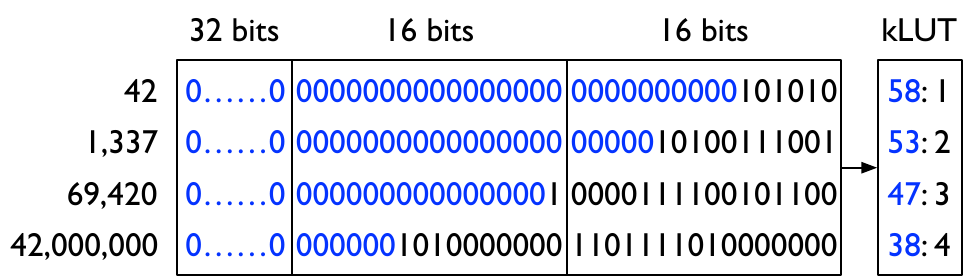}
    \caption{Varint Sizing Example}
    \label{fig:klut}
\end{figure}

To illustrate the algorithm 4, consider a small dataset of four 64-bit integers: 42, 1,337, 69,420, and 42,000,000. Figure~\ref{fig:klut} shows the binary representation of each integer, highlighting the most significant set bit position and the corresponding LUT index and byte size.
The algorithm determines the varint encoding byte size for each integer using the precomputed LUT. For 42, the most significant set bit is at position 58, and the LUT value at index 58 is 1, indicating a 1-byte encoding. Similarly, 1,337 requires 2 bytes (LUT index 53), 69,420 requires 3 bytes (LUT index 47), and 42,000,000 requires 4 bytes (LUT index 38).
The algorithm sums the byte sizes, yielding a total storage requirement of 10 bytes (1 + 2 + 3 + 4) for the dataset. The retrieved byte sizes for each integer are accumulated in the size variable, which represents the total storage requirement for the varint-encoded integers.

The use of a precomputed LUT eliminates the need for runtime calculations and conditional branching, resulting in efficient and fast estimation of varint sizes. The LUT is small enough to fit in the cache, minimizing cache misses and further improving performance. By employing this LUT-based approach, the algorithm achieves rapid and accurate estimation of storage requirements for varint encoding, enabling efficient memory allocation and management in varint-based systems.




\subsection{BMI2-Enhanced Bulk Varint Decoding}

This section enhances varint decoding with BMI2 instructions, advancing beyond the basic methods in Section \ref{basic_varint}.

\vspace{2mm}
\textbf{Mask Configuration}: The first step in our BMI2-based LEB128 varint decoding approach involves the use of a 64-bit mask with the \texttt{\_pext\_u64} instruction for efficient extraction of the most significant bits (MSBs) from byte sequences. The choice of mask configuration is crucial to the performance of our approach, as it directly influences the number of bytes processed simultaneously and the range of values that can be extracted. In SFVInt, we considered three primary mask configurations:
\texttt{0x0000808080808080} for six bytes, \texttt{0x0080808080808080} for seven bytes, and \texttt{0x8080808080808080} for eight bytes. Each of these configurations caters to different ranges of extracted values: 0-63 for six bytes, 0-127 for seven bytes, and 0-255 for eight bytes.

Utilizing \texttt{\_pext\_u64(word, mask)}, we decipher the count and positions of integers within a 64-bit block. For example, using an 8-byte mask (\texttt{0x8080808080808080}) that yields zero indicates eight complete integers in the segment. Initially setting aside carryover or boundary effects for simplicity, we employ \texttt{\_pext\_u64} instructions strategically to isolate individual integers:

{\footnotesize
\begin{verbatim}
         int1 = _pext_u64(word, 0x000000000000007f);
         int2 = _pext_u64(word, 0x0000000000007f00);
         int3 = _pext_u64(word, 0x00000000007f0000);
         int4 = _pext_u64(word, 0x000000007f000000);
         int5 = _pext_u64(word, 0x0000007f00000000);
         int6 = _pext_u64(word, 0x00007f0000000000);
         int7 = _pext_u64(word, 0x007f000000000000);
         int8 = _pext_u64(word, 0x7f00000000000000);
\end{verbatim}
}

For a masked result of 63, indicating three integers spanning six, one and one bytes respectively in the original varint sequence. To extract them efficiently, we apply the \texttt{\_pext\_u64} instruction with specific masks tailored to these lengths:

{\footnotesize
\begin{verbatim}
         int1 = _pext_u64(word, 0x00007f7f7f7f7f7f);
         int2 = _pext_u64(word, 0x007f000000000000);
         int3 = _pext_u64(word, 0x7f00000000000000);
\end{verbatim}
}

This method showcases a high degree of adaptability to various varint structures, which are identified through the mask results obtained from the \texttt{\_pext\_u64} instruction. By leveraging these mask results, our approach can efficiently process the encoded data stream while minimizing the need for conditional byte-by-byte checks, which are a common source of branch mispredictions and pipeline stalls in traditional decoding methods.
This understanding of the mask's role is fundamental to our optimized decoding process. By using the mask results to guide the decoding logic, we can efficiently extract and process the relevant bits from the encoded data stream, resulting in a significant performance improvement over traditional methods.

While longer masks, such as \texttt{0x8080808080808080}, might appear to offer better performance by processing more bytes simultaneously, our empirical findings suggest otherwise. The use of longer masks results in a larger instruction footprint, as the decoding logic needs to handle a greater number of possible varint configurations. This increased complexity can exceed the capacity of the L1 instruction cache, leading to cache misses and performance degradation. To strike an optimal balance between decoding efficiency and CPU resource utilization, we empirically determined that the six-byte mask configuration provides the best performance. This choice aligns with the operational constraints of modern CPU architectures, minimizing the risk of L1 instruction cache spillage while still enabling efficient decoding.


\vspace{2mm}
\textbf{Cross-Boundary Cases}: In addressing cross-boundary cases where integers span two or even three consecutive blocks, we adopt a refined approach crucial for accurate and complete decoding. Our approach employs two key variables: \texttt{shift\_bits} and \texttt{partial\_value}. \texttt{shift\_bits} tracks the necessary bit displacement for integers crossing block boundaries, while \texttt{partial\_value} retains the already decoded segment from a previous block. This system ensures proper alignment and integration of integer segments across blocks, merging \texttt{partial\_value} with new segments to reconstruct complete integers, thereby preserving the accuracy and integrity of the overall decoding process.


Continuing from our previous example where a masked result of 63 suggests two integers with the first spanning six bytes and the second two, we encounter the complexity of \texttt{int1} extending across 64-bit block boundaries. Here, \texttt{partial\_value} retains \texttt{int1}'s initial segment from the preceding block, while \texttt{shift\_bits} tracks the decoded bit count. The subsequent processing extracts \texttt{int1}'s remainder, aligns it using \texttt{shift\_bits}, and merges it with \texttt{partial\_value}, forming the complete integer, as demonstrated below:

{\footnotesize
\begin{verbatim}
        int1 = (_pext_u64(word, 0x00007f7f7f7f7f7f)
                << shift_bits) | partial_value;
        int2 =  _pext_u64(word, 0x007f000000000000);
        int3 =  _pext_u64(word, 0x7f00000000000000);
\end{verbatim}
}

This method effectively handles integers crossing 64-bit block boundaries, ensuring their precise and complete reconstruction. 

\vspace{2mm}
\textbf{Algorithm}: The following presents an advanced varint decoding algorithm optimized for bulk processing with a 6-byte MASK, integrating cross-boundary case handling and fully leveraging BMI2 instructions for improved efficiency and speed. Line 7 employs \texttt{\_pext\_u64} to extract MSBs from a 64-bit word using a defined mask, revealing the varint structure, including the count and distribution of integers. The algorithm then uses this output (\texttt{mval}) in a switch-case to tailor the decoding process for each integer, adapting to their specific encoding patterns within the data stream. To illustrate the decoding process, let's examine three representative cases: case 0, case 45, case 62 and case 63.



 {\footnotesize
\begin{figure}[ht]
\centering
\begin{minipage}{\linewidth}
\begin{verbatim}
Algorithm 5: Optimized Bulk Varint Decoding with BMI2
Input: uint64_t n, BufferPointer buf, T* res (uint32/uint64)
Output: Updated BufferPointer in the data stream
1:  mask_length = 6; mask = 0x0000808080808080;
2:  shift_bits = 0; pt_val = partial_value = 0;
3:  buf -= mask_length;
4:  while n >= 8 do
5:    buf += mask_length;
6:    word = *(const uint64_t*)(buf);
7:    mval = _pext_u64(word, mask);
8:    switch (mval)
9:      case 0: 
10:         cu_val = _pext_u64(word, 0x000000000000007f);
11:         *res++ = (cu_val << shift_bits) | pt_val;
12:         *res++ = _pext_u64(word, 0x0000000000007f00);
13:         *res++ = _pext_u64(word, 0x00000000007f0000);
14:         *res++ = _pext_u64(word, 0x000000007f000000);
15:         *res++ = _pext_u64(word, 0x0000007f00000000);
16:         *res++ = _pext_u64(word, 0x00007f0000000000);
17:         shift_bits = 0; pt_val = 0; n -= 6;
18:      case 45:
19:         cu_val = _pext_u64(word, 0x0000000000007f7f);
20:         *res++ = (cu_val << shift_bits) | pt_val;
21:         *res++ = _pext_u64(word, 0x0000007f7f7f0000);
22:         pt_val = _pext_u64(word, 0x00007f0000000000);
23:         shift_bits = 7; n -= 2;
24:      case 62:
25:         cu_val = _pext_u64(word, 0x000000000000007f);
26:         *res++ = (cu_val << shift_bits) | pt_val;
27:         pt_val = _pext_u64(word, 0x00007f7f7f7f7f00);
28:         shift_bits = 35; n -= 1;
29:      case 63:
30:         pt_val |= _pext_u64(word, 0x00007f7f7f7f7f7f)
31:                << shift_bits;
32:         shift_bits += 42;
33:      case ... // Other cases omitted for brevity
34: end while
\end{verbatim}
\end{minipage}
\caption{BMI2-enhanced Bulk Decoding.}
\label{fig:bulk_decoding}
\end{figure}
}

Case 0 (lines 10-17) handles the scenario where the current 64-bit word contains six complete single-byte integers. When \texttt{mval} is 0, it indicates that all six bytes in the word represent individual integers. The algorithm proceeds to extract each integer using \texttt{\_pext\_u64} with carefully crafted masks targeting their respective positions within the word. The extracted integers are then stored in the output array \texttt{res}, and the \texttt{shift\_bits} and \texttt{pt\_val} variables are reset to prepare for the next iteration. Finally, the count of remaining integers to decode (n) is decremented by 6 (line 17).

Case 45 (lines 19-23) handles the scenario where the current 64-bit word contains two 2-byte integers. When \texttt{mval} is 45 (binary representation: \texttt{1\textbf{011}01}), it indicates that the first integer spans across the first two bytes of the word, while the second integer spans across the third and fifth bytes. The algorithm extracts these integers using \texttt{\_pext\_u64} with masks \texttt{0x0000000000007f7f} and \texttt{0x0000007f7f7f0000}, respectively (lines 19-21). The \texttt{pt\_val} variable is updated with any remaining partial value (line 22), and the \texttt{shift\_bits} variable is set to 7 to align the decoding process for the next iteration. Finally, the count of remaining integers (\texttt{n}) is decremented by 2 (line 23).

Case 62 (lines 25-28) demonstrates the algorithm's ability to handle integers that span across multiple 64-bit words. When \texttt{mval} is 62 (binary representation: 111110) under a 6-byte mask, it signifies that an integer ends within the current word, with its remaining bits located in the preceding word. The algorithm extracts the relevant bits from the current word using \texttt{\_pext\_u64} and merges them with the partial value (\texttt{pt\_val}) from the previous word to reconstruct the complete integer (lines 26). Moreover, the presence of five leading 1s (11111) in \texttt{mval} indicates an incomplete integer extending into the next word. This partial integer is extracted and stored in \texttt{pt\_val} for subsequent decoding, and the \texttt{shift\_bits} variable is updated to 35 (the five 7-byte sequences) to align the decoding process for the next iteration. This decrement in \texttt{n} by 1 (line 28) ensures the algorithm's seamless management of cross-boundary integers, maintaining decoding continuity and precision.

Case 63, depicted in lines 30-32, addresses a binary representation of \texttt{111111}, signaling an integer spanning three data blocks, with the current block as its mid-segment. This specific case requires extracting six bytes (42 bits) via \texttt{\_pext\_u64}, appending this data to the partial value to maintain integer continuity across blocks.

In addition to the representative cases discussed above, our algorithm includes a comprehensive set of cases to handle various varint encoding patterns efficiently. These cases, omitted for brevity in the pseudocode (line 33), cover a wide range of scenarios, ensuring optimized decoding performance for different integer distributions.

The seamless handling of cross-boundary integers is a key strength of our approach. By maintaining state variables like \texttt{shift\_bits} and \texttt{pt\_val}, we ensure the continuity and precision of the decoding process across word boundaries. The decrement of n by the appropriate count in each case guarantees that the algorithm keeps track of the remaining integers to decode, maintaining the correctness of the output.
Through this case analysis, we showcase the adaptability and efficiency of our BMI2-based varint decoding algorithm. By leveraging the power of the \texttt{\_pext\_u64} instruction and carefully designing the decoding logic for each case, we can handle a wide range of varint encoding patterns with minimal branching and maximum performance.




\section{Performance Evaluation}
\begin{figure*}
    \centering
    \begin{minipage}[b]{0.578\textwidth}
        \centering
        \includegraphics[width=\textwidth]{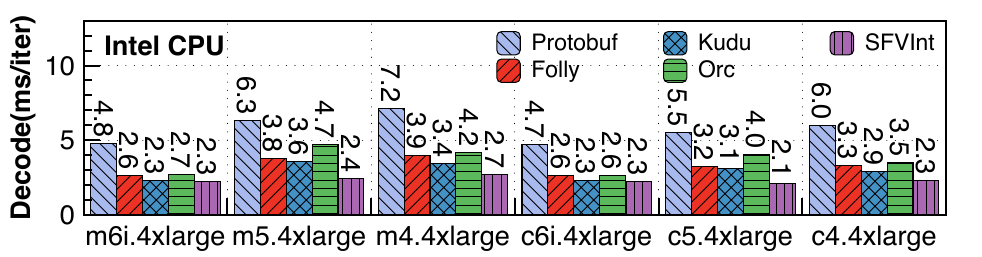}
    \end{minipage}
    \begin{minipage}[b]{0.417\textwidth}
        \centering
        \includegraphics[width=\textwidth]{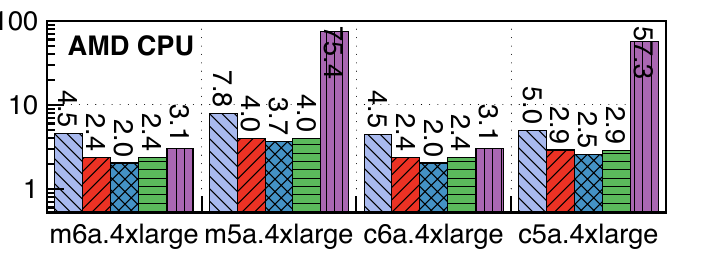}
    \end{minipage}
    \caption{Workload 1 (W1): uniform distribution.}
        \label{w1}
\end{figure*}

\begin{figure*}
    \centering
    \begin{minipage}[b]{0.578\textwidth}
        \centering
        \includegraphics[width=\textwidth]{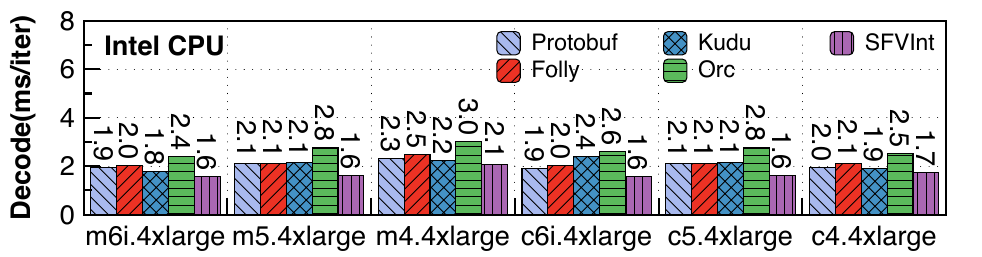}
    \end{minipage}
    \begin{minipage}[b]{0.417\textwidth}
        \centering
        \includegraphics[width=\textwidth]{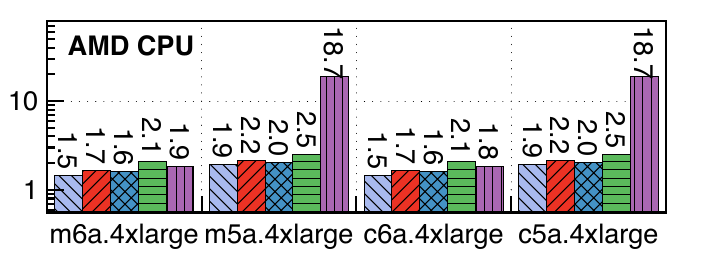}
    \end{minipage}
    \caption{W2: varint byte distribution — {\bf 1 byte: 90.08\%, 2 bytes: 4.63\%, 3 bytes: 3.22\%, 4 bytes: 1.20\%, 5 bytes: 0.88\%}.}
    \label{w2}
\end{figure*}

\begin{figure*}
    \centering
    \begin{minipage}[b]{0.578\textwidth}
        \centering
        \includegraphics[width=\textwidth]{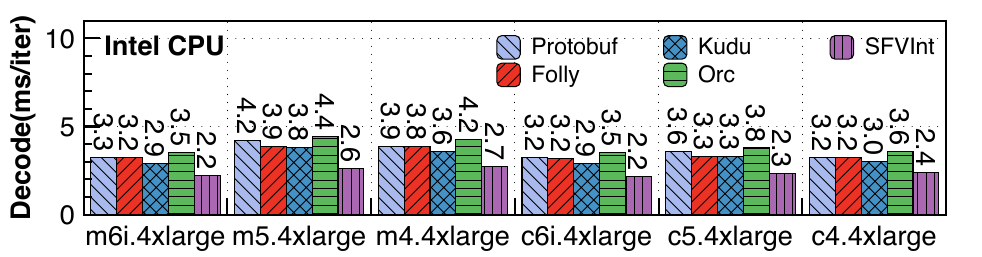}
    \end{minipage}
    \begin{minipage}[b]{0.417\textwidth}
        \centering
        \includegraphics[width=\textwidth]{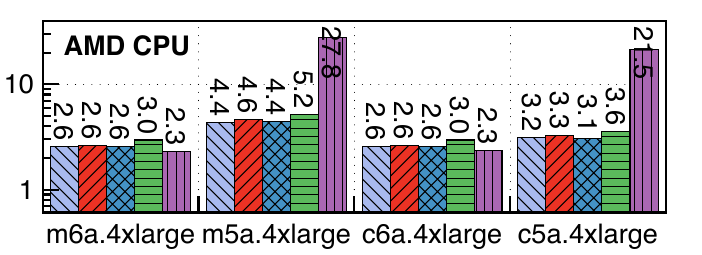}
    \end{minipage}
    \caption{W3: varint byte distribution — {\bf 1 byte: 81.22\%, 2 bytes: 7.31\%, 3 bytes: 6.16\%, 4 bytes: 4.20\%, 5 bytes: 1.10\%}.}
    \label{w3}
\end{figure*}

\begin{figure*}
    \centering
    \begin{minipage}[b]{0.578\textwidth}
        \centering
        \includegraphics[width=\textwidth]{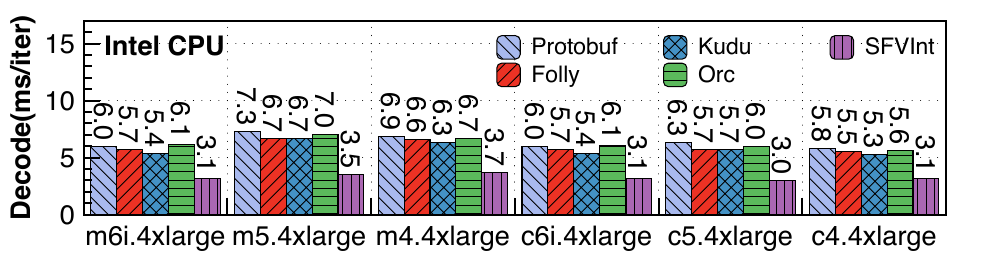}
    \end{minipage}
    \begin{minipage}[b]{0.417\textwidth}
        \centering
        \includegraphics[width=\textwidth]{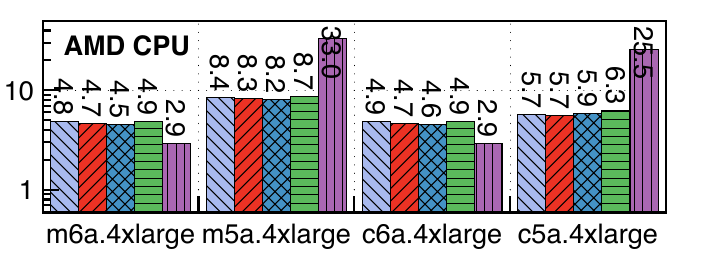}
    \end{minipage}
    \caption{W4: varint byte distribution — {\bf 1 byte: 72.13\%, 2 bytes: 12.31\%, 3 bytes: 8.53\%, 4 bytes: 5.31\%, 5 bytes: 1.72\%}.}
    \label{w4}
\end{figure*}


In our comprehensive evaluation of the BMI2-enhanced varint decoding technique, we utilized a diverse set of AWS EC2 instances equipped with a range of Intel and AMD CPU architectures (refer to Table~\ref{tab:ec2_cpu_arch}). This heterogeneous testing environment allowed us to assess the performance and versatility of our approach across different processor generations and microarchitectures. To thoroughly evaluate the efficiency of our technique, we generated datasets encoded as varints following two distinct distributions: uniform and skewed. The uniform distribution dataset served as a stable benchmark, providing a balanced representation of integer values across the entire range. This dataset enabled us to assess the performance of our approach under a consistent and predictable workload. On the other hand, the skewed distribution dataset was specifically designed to mirror real-world varint patterns commonly encountered in various applications and data formats. This dataset exhibited a bias towards smaller integers, with a higher frequency of 1-byte and 2-byte LEBs. By incorporating a skewed distribution, we aimed to evaluate the performance of our technique in scenarios that closely resemble practical use cases.


We employed four distinct workloads for evaluation: W1 featured uniformly distributed 32-bit integers as a benchmark; W2 used real-world LEB lengths from the WebAssembly build suite~\cite{webassembly}, providing practical insights; and Workloads W3 and W4 derived from our proprietary systems' data distributions, offering a realistic view of their application. Notably, W2 to W4 exhibited a skew towards 1-byte integers, aligning with typical data trends observed in practical scenarios. The specific distributions for these workloads are detailed in the figure captions of Figures \ref{w1} to \ref{w4}.

Our evaluation compares our BMI2-based decoding technique against established varint encoding solutions from Apache Kudu (a columnar storage system)\cite{lipcon2015kudu}, Google Protobuf (a serde library)\cite{protobuf}, Apache ORC (a columnar format)\cite{orc}, and Facebook Folly (a C++ std lib)\cite{folly}, which all support 32-bit and 64-bit integer encoding. This benchmark situates our approach within the current industry practices\footnote{Our assessment directly interacted with the varint decoding interfaces of these libraries, ensuring no additional overhead is incurred.}, focusing on in-memory decoding performance given the widespread use of the LEB128 algorithm. Performance gains were assessed in a single-threaded environment to attribute improvements directly to our method.


\begin{table}
\small
\centering
\begin{tabular}{|l|l|}
\hline
\textbf{EC2} & \textbf{CPU Architecture} \\ \hline
m6i & Intel Xeon Ice Lake 8375C \\ \hline
m5 & Intel Xeon Skylake 8175M / Cascade Lake 8259CL \\ \hline
m4 & Intel Xeon Broadwell E5-2686 / Haswell E5-2676 \\ \hline
c6i & Intel Xeon Ice Lake 8375C \\ \hline
c5 & Intel Xeon Cascade Lake \\ \hline
c4 & Intel Xeon Haswell E5-2666 v3 \\ \hline\hline
m6a & 3rd Gen AMD EPYC (Milan) \\ \hline
c6a & 3rd Gen AMD EPYC (Milan) \\ \hline
m5a & 2nd Gen AMD EPYC 7571 \\ \hline
c5a & 2nd Gen AMD EPYC 7R32 \\ \hline
\end{tabular}
\vspace{3mm}
\caption{AWS EC2 Types and CPU Architectures}
\label{tab:ec2_cpu_arch}
\end{table}

\subsection{Intel CPU Microarchitectures}

In Figures \ref{w1}-\ref{w4}, the left panel showcases the varint decoding performance across six EC2 instance types featuring Intel CPUs, detailed in Table~\ref{tab:ec2_cpu_arch}. Each figure within the panel includes benchmarks for each EC2 instance type, with the y-axis representing the time required per iteration (in milliseconds), where each iteration involves decoding one million integers generated according to a specific distribution. Here, \texttt{SFVInt} consistently excels, surpassing all other systems in decoding speed across every instance type and under various distributions. For example, in W1 (Figure~\ref{w1}), \texttt{SFVInt} demonstrates a significant performance edge, nearly doubling the speed of Protobuf. Similarly, in W4 (Figure~\ref{w4}), \texttt{SFVInt} outperforms all other systems by nearly a factor of two.

The varint byte distribution evolves from W2 to W4, indicating a transition to a more varied spread of integer sizes. In W2 (Figure~\ref{w2}), the distribution heavily favors 1-byte integers, constituting 90.08\% of the data, with a minor representation of 2-byte at 4.63\%, and even less for 3-byte (3.22\%), 4-byte (1.20\%), and 5-byte integers (0.88\%). Progressing to W4, there's a noticeable decrease in 1-byte integers to 72.13\%, with corresponding increases across the board—2-byte integers rise to 12.31\%, 3-byte to 8.53\%, 4-byte to 5.31\%, and 5-byte to 1.72\%. This shift implies an increase in the number of bytes subject to the decoding process, logically resulting in prolonged iteration times as the byte distribution broadens. Specifically, on the \texttt{m6i.4xlarge} instance, the iteration time for \texttt{SFVInt} extends from 1.6ms in W2, to 2.2ms in W3, ultimately reaching 3.1ms in W4. Such an increment reflects the additional computational overhead necessitated by the decoding of a higher proportion of multi-byte integers.

On the other hand, the performance gap between SFVInt and other systems widens most notably in W4. Taking the c6i.4xlarge instance as an example, SFVInt achieves a 2x speed increase compared to Protobuf in W4,  while in W3 and W2, the gains are 45\% and 19\%, respectively. This indicates that \texttt{SFVInt}'s BMI2-fortified decoding mechanism gains a significant edge when dealing with varints exceeding the single-byte range. This is because with shorter byte-length encodings, our approach requires a sequence of serial instructions similar to traditional byte-by-byte processing. This is exemplified by case 0 (lines 10 - 17) in Figure~\ref{fig:bulk_decoding}, which involves a greater number of sequential executions than case 63 (lines 44 - 46), hence, SFVInt's relatively subdued enhancement for 1-byte integers reflects this similarity to conventional decoding approaches.

\subsection{AMD CPU Microarchitectures}


In our analysis detailed in Figures \ref{w1}-\ref{w4}, we observe that \texttt{SFVInt} demonstrates up to 40\% faster varint decoding on 3rd generation AMD EPYC processors (\texttt{m6a} and \texttt{c6a}), in contrast to the slower performance on 2nd generation CPUs (\texttt{m5a} and \texttt{c5a}). This performance disparity is consistent with the known emulation overhead of PEXT/PDEP instructions on AMD platforms, which can significantly lag behind Intel's native execution~\cite{kullberg2017bmi2,reddit2017bmi2}. Studies indicate that on AMD hardware, these instructions' latency can vary widely, from 18 to approximately 300 cycles~\cite{amd-bmi-latency}. The improved results with newer AMD generations suggest advancements in BMI2 instruction handling. For enhanced consistency across CPU architectures, we can incorporate a dynamic selection mechanism between BMI2-accelerated and standard decoding or employs ZP7~\cite{zp7}, a branchless alternative for AMD processors.



\section{Related Works}

{\bf Varint Formats.}
In byte-oriented integer compression, various techniques aim to improve decoding efficiency. Traditional LEB128, using a 7-bit structure, lacks SIMD optimization. In contrast, Google's VARINT-GB~\cite{jeff37797} offers an alternative by encoding a constant block of integers with a unified control byte, streamlining decoding for data with variable lengths. The patented VARINT-G8IU~\cite{10.1145/2063576.2063627} further advances this innovation, employing SIMD to process variable counts of integers within a fixed byte structure, controlled by an initial byte. Stream VByte~\cite{Lemire_2018} segregates control and data bytes into separate streams to fully leverage SIMD's parallel processing strengths. However, these techniques, despite their SIMD-enabled efficiency gains, diverge from the prevalent LEB128 standard, limiting their seamless integration into existing infrastructures. Conversely, the Masked VByte method~\cite{plaisance2017vectorized} retains the LEB128 form while harnessing SIMD for improved performance, yet it is not without its drawbacks—namely, a non-generic and maintenance-intensive codebase exceeding 2300 lines, tailored specifically for 32-bit unsigned integers. Our proposed SFVInt is a simple, elegant, fast, and generic solution that effectively addresses the limitations present in earlier methods.

\vspace{2mm}
{\bf \noindent BMI in DBMS.}
The pursuit of SIMD optimizations for data scanning operations and operator vectorization has been prominent in data management research~\cite{BitWeaving, light_compression_fast_scan,regular_expr_match_simd,bloom_filter_simd}. 
Kudu was seminal in applying BMI2 for data scan filtering~\cite{lipcon2015kudu,kudu-bmi}, a similar idea extended by Parquet-Select~\cite{yinan2023}, which utilizes BMI2 for optimized selection pushdown in columnar storage. 

During our investigation of next-generation columnar store for machine learning workloads~\cite{liao2024bullion}, we found discussions on Google Protobuf regarding the use of BMI2 for accelerating varint decoding~\cite{zhaozhou,andrei}. The Protobuf team expressed concerns about the performance of BMI2 on AMD processors. Motivated by these discussions, we developed SFVInt and conducted a comprehensive analysis of the effectiveness of BMI2 on both Intel and AMD platforms, addressing the performance considerations raised by the community.

\section{Conclusions}

In this paper, we presented SFVInt, a simple, fast, and generic approach for decoding variable-length integers using BMI2 instructions. Through extensive evaluation across diverse CPU architectures, SFVInt consistently demonstrated up to a 2x increase in decoding speed compared to traditional methods, addressing a critical bottleneck in data-intensive applications. As data continues to grow in volume and complexity, SFVInt represents a step forward in offering a high-performance solution for varint decoding that can greatly benefit a wide range of applications.


\newpage


\bibliographystyle{ACM-Reference-Format}
\bibliography{sample-base}










\end{document}